\documentclass[aps,prd,twocolumn,tightenlines,superscriptaddress,amsmath,amssymb]{revtex4}
\usepackage{graphicx} %
\usepackage{epstopdf} %
\usepackage{dcolumn} %
\usepackage{xcolor} %
\usepackage{amsmath,amssymb} %
\usepackage{hyperref} %
\usepackage[utf8]{inputenc} %
\usepackage[english]{babel} %
\usepackage[mathlines]{lineno} %
\usepackage{blindtext} %
\usepackage{ifthen} %
\usepackage{orcidlink} %

\usepackage[noabbrev,capitalise]{cleveref} %

\usepackage{afterpage} %
\usepackage{rotating,booktabs,multirow} %

\usepackage[percent]{overpic}

\graphicspath{{figures/}} %

\newboolean{articletitles}
\setboolean{articletitles}{true} %

\newcommand*\patchAmsMathEnvironmentForLineno[1]{%
\expandafter\let\csname old#1\expandafter\endcsname\csname #1\endcsname
\expandafter\let\csname oldend#1\expandafter\endcsname\csname
end#1\endcsname
 \renewenvironment{#1}%
   {\linenomath\csname old#1\endcsname}%
   {\csname oldend#1\endcsname\endlinenomath}%
}
\newcommand*\patchBothAmsMathEnvironmentsForLineno[1]{%
  \patchAmsMathEnvironmentForLineno{#1}%
  \patchAmsMathEnvironmentForLineno{#1*}%
}
\AtBeginDocument{%
\patchBothAmsMathEnvironmentsForLineno{equation}%
\patchBothAmsMathEnvironmentsForLineno{align}%
\patchBothAmsMathEnvironmentsForLineno{flalign}%
\patchBothAmsMathEnvironmentsForLineno{alignat}%
\patchBothAmsMathEnvironmentsForLineno{gather}%
\patchBothAmsMathEnvironmentsForLineno{multline}%
\patchBothAmsMathEnvironmentsForLineno{eqnarray}%
}

\RequirePackage{xspace}
\RequirePackage{relsize}

\def\belletwo{\mbox{Belle~II}\xspace}

\def\babar{\mbox{\slshape B\kern-0.1em{\smaller A}\kern-0.1em
    B\kern-0.1em{\smaller A\kern-0.2em R}}\xspace}

\def\Pmu         {\ensuremath{\mu}\xspace}

\def\Ppi         {\ensuremath{\pi}\xspace}

\mathchardef\PDelta="7101
\mathchardef\PXi="7104
\mathchardef\PLambda="7103
\mathchardef\PSigma="7106
\mathchardef\POmega="710A
\mathchardef\PUpsilon="7107
                 
\def\PB      {\ensuremath{B}\xspace}                 
                 
\def\PD      {\ensuremath{D}\xspace}

\def\PK      {\ensuremath{K}\xspace}

\def\Pc      {\ensuremath{c}\xspace}                 
                 
\def\Pe      {\ensuremath{e}\xspace}

\def\Pi      {\ensuremath{i}\xspace}

\def\Pq      {\ensuremath{q}\xspace}                 
                 
\def\Ps      {\ensuremath{s}\xspace}

\def\epem       {\ensuremath{\Pe^+\Pe^-}\xspace}

\def\mup        {\ensuremath{\Pmu^+}\xspace}

\def\quark     {\ensuremath{\Pq}\xspace}
\def\quarkbar  {\ensuremath{\overline \quark}\xspace}
\def\qqbar     {\ensuremath{\quark\quarkbar}\xspace}

\def\squark    {\ensuremath{\Ps}\xspace}

\def\cquark    {\ensuremath{\Pc}\xspace}
\def\cquarkbar {\ensuremath{\overline \cquark}\xspace}
\def\ccbar     {\ensuremath{\cquark\cquarkbar}\xspace}

\def\pion  {\ensuremath{\Ppi}\xspace}
\def\piz   {\ensuremath{\pion^0}\xspace}

\def\pip   {\ensuremath{\pion^+}\xspace}
\def\pim   {\ensuremath{\pion^-}\xspace}

\def\kaon  {\ensuremath{\PK}\xspace}
\def\Kbar  {\kern 0.2em\overline{\kern -0.2em \PK}{}\xspace}%

\def\Kz    {\ensuremath{\kaon^0}\xspace}
\def\Kzb   {\ensuremath{\Kbar^0}\xspace}
\def\KzKzb {\ensuremath{\Kz \kern -0.16em \Kzb}\xspace}
\def\Kp    {\ensuremath{\kaon^+}\xspace}
\def\Km    {\ensuremath{\kaon^-}\xspace}
\def\Kpm   {\ensuremath{\kaon^\pm}\xspace}

\def\KpKm  {\ensuremath{\Kp \kern -0.16em \Km}\xspace}
\def\KS    {\ensuremath{\kaon^0_{\rm\scriptscriptstyle S}}\xspace} 
\def\KL    {\ensuremath{\kaon^0_{\rm\scriptscriptstyle L}}\xspace}

\def\D       {\ensuremath{\PD}\xspace}
\def\Dbar    {\kern 0.2em\overline{\kern -0.2em \PD}{}\xspace}%

\def\Dz      {\ensuremath{\D^0}\xspace}
\def\Dzb     {\ensuremath{\Dbar^0}\xspace}
\def\DzDzb   {\ensuremath{\Dz {\kern -0.16em \Dzb}}\xspace}
\def\Dp      {\ensuremath{\D^+}\xspace}
\def\Dm      {\ensuremath{\D^-}\xspace}

\def\DpDm    {\ensuremath{\Dp {\kern -0.16em \Dm}}\xspace}

\def\Dstarp  {\ensuremath{\D^{*+}}\xspace}

\def\Dsp     {\ensuremath{\D^+_\squark}\xspace}

\def\B       {\ensuremath{\PB}\xspace}
\def\Bbar    {\ensuremath{\kern 0.18em\overline{\kern -0.18em \PB}{}}\xspace}%

\def\Bz      {\ensuremath{\B^0}\xspace}

\def\Bu      {\ensuremath{\B^+}\xspace}

\def\Bp      {\ensuremath{\Bu}\xspace}

\def\Bpm     {\ensuremath{\B^\pm}\xspace}

\def\Bs      {\ensuremath{\B^0_\squark}\xspace}

\def\Y#1S{\ensuremath{\PUpsilon{(#1S)}}\xspace}%

\def\L {\ensuremath{\PLambda}\xspace}
\def\Lbar {\ensuremath{\kern 0.1em\overline{\kern -0.1em\PLambda}}\xspace}

\def\Lz      {\ensuremath{\L^0}\xspace}

\def\Lc      {\ensuremath{\L^+_\cquark}\xspace}

\def\to                 {\ensuremath{\rightarrow}\xspace}

\def\CP                {\ensuremath{C\!P}\xspace}

\newcommand{\tev}{\ensuremath{\mathrm{\,Te\kern -0.1em V}}\xspace}
\newcommand{\gev}{\ensuremath{\mathrm{\,Ge\kern -0.1em V}}\xspace}
\newcommand{\mev}{\ensuremath{\mathrm{\,Me\kern -0.1em V}}\xspace}
\newcommand{\kev}{\ensuremath{\mathrm{\,ke\kern -0.1em V}}\xspace}
\newcommand{\ev}{\ensuremath{\mathrm{\,e\kern -0.1em V}}\xspace}
\newcommand{\gevc}{\ensuremath{{\mathrm{\,Ge\kern -0.1em V\!/}c}}\xspace}
\newcommand{\mevc}{\ensuremath{{\mathrm{\,Me\kern -0.1em V\!/}c}}\xspace}
\newcommand{\gevcc}{\ensuremath{{\mathrm{\,Ge\kern -0.1em V\!/}c^2}}\xspace}
\newcommand{\gevgevcccc}{\ensuremath{{\mathrm{\,Ge\kern -0.1em V^2\!/}c^4}}\xspace}
\newcommand{\mevcc}{\ensuremath{{\mathrm{\,Me\kern -0.1em V\!/}c^2}}\xspace}

\def\cm   {\ensuremath{\rm \,cm}\xspace}

\def\invfb   {\ensuremath{\mbox{\,fb}^{-1}}\xspace}

\newcommand{\chisq}{\ensuremath{\chi^2}\xspace}

\def\gsim{{~\raise.15em\hbox{$>$}\kern-.85em
          \lower.35em\hbox{$\sim$}~}\xspace}
\def\lsim{{~\raise.15em\hbox{$<$}\kern-.85em
          \lower.35em\hbox{$\sim$}~}\xspace}

\newcommand{\ie}{\mbox{\itshape i.e.}}

\newcommand{\Acp}{\ensuremath{A_{\CP}}\xspace}

\newcommand{\resValueTagg}{-3.9}
\newcommand{\resStatTagg}{1.8}
\newcommand{\resSystTagg}{0.2}

\newcommand{\resValueNull}{-1.1}
\newcommand{\resStatNull}{1.0}
\newcommand{\resSystNull}{0.1}

\newcommand{\resValueComb}{-1.8}
\newcommand{\resStatComb}{0.9}
\newcommand{\resSystComb}{0.1}

\usepackage{ifthen} %
\newboolean{wordcount}
\setboolean{wordcount}{false} %

\ifthenelse{\boolean{wordcount}}%
{\usepackage{bibentry} %
 \usepackage{comment} %
 \excludecomment{align*} %
 \excludecomment{align} %
 \excludecomment{equation*} %
 \excludecomment{equation} %
 \excludecomment{eqnarray*} %
 \excludecomment{eqnarray} %
 \excludecomment{acknowledgments} %
 \def\maketitle{} %
}{
}

\makeatletter%
\begin{document}

\ifthenelse{\boolean{wordcount}}{}
{
\includegraphics[width=3cm]{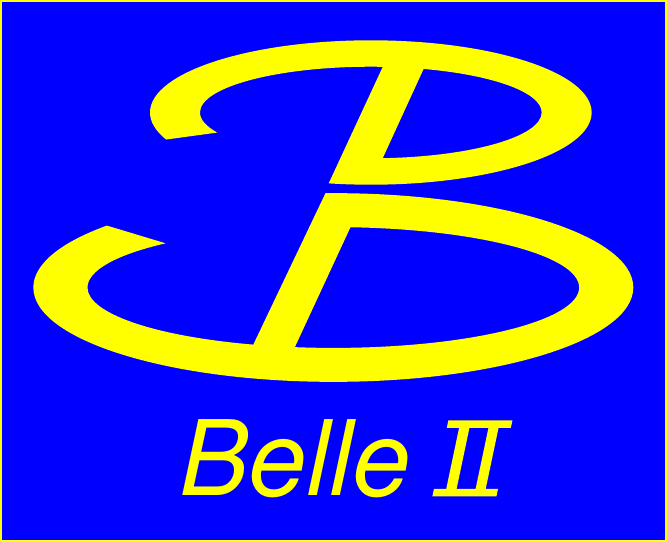}\vspace*{-1.9cm}
\begin{flushright}
Belle II Preprint 2025-012\\
KEK Preprint 2025-10
\end{flushright}\vspace{1.5cm}
}

\title{%
%
{\LARGE\bfseries\boldmath Measurement of the \CP asymmetry in $\Dp\to\pip\piz$ decays at Belle II}
 %
}
%
%
%
%
%
%
  \author{I.~Adachi\,\orcidlink{0000-0003-2287-0173}} %
  \author{L.~Aggarwal\,\orcidlink{0000-0002-0909-7537}} %
  \author{H.~Ahmed\,\orcidlink{0000-0003-3976-7498}} %
  \author{H.~Aihara\,\orcidlink{0000-0002-1907-5964}} %
  \author{N.~Akopov\,\orcidlink{0000-0002-4425-2096}} %
  \author{S.~Alghamdi\,\orcidlink{0000-0001-7609-112X}} %
  \author{M.~Alhakami\,\orcidlink{0000-0002-2234-8628}} %
  \author{A.~Aloisio\,\orcidlink{0000-0002-3883-6693}} %
  \author{K.~Amos\,\orcidlink{0000-0003-1757-5620}} %
  \author{M.~Angelsmark\,\orcidlink{0000-0003-4745-1020}} %
% ...
%
\collaboration{The Belle II Collaboration}
 \begin{abstract}
%
%
\noindent We measure the \CP asymmetry in $\Dp\to\pip\piz$ decays reconstructed in \epem collisions at the \belletwo experiment using a dataset corresponding to an integrated luminosity of 428\invfb. A control sample of $\Dp\to\pip\KS$ decays is used to correct for detection and production asymmetries. The result, $\Acp(D^+ \to \pi^+\pi^0) =(\resValueComb\pm\resStatComb\pm\resSystComb)\%$, where the first uncertainty is statistical and the second systematic, is the most precise determination to date. It agrees with the prediction of \CP symmetry from the standard model, and with results of previous measurements. 
 \end{abstract}

\maketitle

%
%
%
In the standard model of particle physics, charge-parity (\CP) violation arises from the complex phase of the Cabibbo-Kobayashi-Maskawa matrix~\cite{Cabibbo:1963yz,Kobayashi:1973fv} that governs the weak interactions of quarks. Experimental efforts over several decades have observed \CP violation in processes involving \Kz, \Bp, \Bz, and \Bs mesons with results consistent with standard model predictions~\cite{Christenson:1964fg,KTeV:1999kad,NA48:1999szy,BaBar:2002kla,Belle:2002ghj,Aubert:2004qm,Chao:2004mn,LHCb-PAPER-2012-001,LHCb-PAPER-2013-018}, but at a level that is insufficient to explain the matter-antimatter asymmetry of the Universe. Measurements in the charm sector have only recently achieved a sufficient level of precision to be sensitive to \CP violation in charm transitions, which is suppressed due to the Glashow-Iliopoulos-Maiani mechanism~\cite{Glashow:1970gm} and the small size of the Cabibbo-Kobayashi-Maskawa matrix element $|V_{cb}|$~\cite{Golden:1989qx,Buccella:1994nf,Bianco:2003vb,Grossman:2006jg,Artuso:2008vf}. The only observation of \CP violation in charm comes from a single measurement of the difference between the time-integrated \CP asymmetries of $D^0\to\Kp\Km$ and $D^0\to\pip\pim$ decays~\cite{Aaij:2019kcg}, along with strong evidence that \CP violation occurs mainly in the direct decay $\Dz\to\pip\pim$~\cite{LHCb:2022lry}. (Charge-conjugate modes are implied throughout, unless stated otherwise.) Nonperturbative QCD effects make it difficult to determine whether the measured \CP asymmetry is consistent with standard model expectations~\cite{Chala:2019fdb,Dery:2019ysp,Calibbi:2019bay,Grossman:2019xcj,Cheng:2019ggx,Buras:2021rdg,Schacht:2021jaz,Bediaga:2022sxw,Bause:2022jes,Schacht:2022kuj,Pich:2023kim,Gavrilova:2023fzy}. Flavor and isospin symmetries can be used to relate measurements from different decay modes, helping to constrain nonperturbative QCD effects and identify possible beyond-standard-model contributions~\cite{Grossman:2012,Hiller:2012xm,Buccella:2019kpn}.

The $D^+ \to \pi^+ \pi^0$ decay is of particular interest. To a good approximation, the standard model generates direct \CP violation in the isospin-related $\Dz\to\pip\pim$ decay through the interference of a leading tree-level amplitude and a suppressed QCD-loop amplitude that changes isospin by half a unit~\cite{Grossman:2012}. Unlike $\pip\pim$, the $\pip\piz$ final state has isospin $I=2$ and can be reached from the $I=1/2$ initial state only via a $\Delta I=3/2$ transition. In the absence of interference with a second amplitude, no \CP violation is expected in $\Dp\to\pip\piz$ decays. Therefore, any observation at the current level of sensitivity will unambiguously indicate physics beyond the standard model~\cite{Grossman:2012,Buccella:1992sg}. Measurements of the \CP asymmetry in $\Dp\to\pip\piz$, which is defined as
\begin{equation}
\Acp(\Dp\to\pip\piz) = \frac{\Gamma(\Dp\to\pip\piz)-\Gamma(\Dm\to\pim\piz)}{\Gamma(\Dp\to\pip\piz)+\Gamma(\Dm\to\pim\piz)}\,,
\end{equation}
with $\Gamma$ being the partial decay width, are all consistent with zero~\cite{CLEO:2009fiz,Belle:2017tho,LHCb:2021rou}. The most precise result, $(-1.3\pm0.9\pm0.6)\%$, where the first uncertainty is statistical and the second systematic, is from the LHCb experiment~\cite{LHCb:2021rou}. It is based on a sample of $28.7\times10^{3}$ $\Dp\to\pip\piz(\to e^+e^-\gamma)$ decays reconstructed in a 9\invfb sample of $pp$ collisions. Reconstructing the neutral pion in the $e^+e^-\gamma$ final state %
enables the determination of the displaced \Dp decay vertex, which helps to suppress background due to particles produced in the primary $pp$ interaction. The Belle result, $(2.3\pm1.2\pm0.2)\%$, is based on $108\times10^{3}$ $\Dp\to\pip\piz(\to\gamma\gamma)$ decays reconstructed in a 921\invfb sample of \epem collisions~\cite{Belle:2017tho}. Despite having a larger signal yield than LHCb's, the Belle sample has substantially larger background from misreconstructed \piz candidates and neutral pions originating from unrelated processes, which degrades the measurement precision. 

In this paper, we present a measurement of the \CP asymmetry in $\Dp\to\pip\piz$ decays using $\epem\to\ccbar$ data collected by Belle II, which have an integrated luminosity of 428\invfb~\cite{lumi}. By employing an improved reconstruction and selection of the $\Dp\to\pip\piz(\to\gamma\gamma)$ decay, we achieve substantially better signal purity and precision compared to both Belle and LHCb. The raw asymmetry between the observed yields of \Dp and \Dm candidates, 
\begin{equation}
A^{\pi^+ \pi^0} = \frac{N(D^+ \to \pi^+ \pi^0) - N(D^- \to \pi^- \pi^0)}{N(D^+ \to \pi^+ \pi^0) + N(D^- \to \pi^- \pi^0)} \,,
\end{equation}
is determined using a fit to the $\pip\piz$ mass distribution. The raw asymmetry can be approximated as the linear combination of the \CP asymmetry, the forward-backward asymmetric production of \Dp and \Dm mesons in $\epem\to\ccbar$ events ($A_{P}^D$)~\cite{Berends:1973fd,Brown:1973ji,Cashmore:1985vp}, and the instrumental asymmetry in detection and reconstruction of $\pi^+$ and $\pi^-$ mesons ($A_{\varepsilon}^{\pi^+}$),
\begin{equation}
A^{\pip\piz} = \Acp(\Dp\to\pip\piz) + A^D_{P} + A_{\varepsilon}^{\pi^+}\,.
\label{eq:Araw:signal}
\end{equation}
We correct for the latter two terms using an abundant control sample of Cabibbo-favored $\Dp\to\pi^+\KS$ decays, where no direct \CP violation is expected. The raw asymmetry of $\Dp\to\pip\KS$ decays,
\begin{equation}\label{eq:Araw:control}
A^{\pi^+\KS}  = A^D_{P} + A_{\varepsilon}^{\pi^+} + A^{\Kzb}\,,
\end{equation}
receives contributions from the same production and detection asymmetries that affect the signal decays, and from effects due to \CP violation and detection of the neutral kaon ($A^{\Kzb}$). The latter can be estimated using the known time evolution of the \Kz-\Kzb system including regeneration effects due to the interactions with the detector material. Thus, the \CP asymmetry of interest is derived as
\begin{equation}\label{eq:Acp}
\Acp(D^+ \to \pi^+ \pi^0) = A^{\pi^+\pi^0} - A^{\pi^+\KS} + A^{\Kzb}\,.
\end{equation}
Variations of $A^D_P$ or $A_{\varepsilon}^{\pi^+}$ due to kinematic differences between signal and control modes are investigated and treated as a source of systematic uncertainty. To improve sensitivity, we categorize signal and control \Dp decays into two classes, named tagged and null tag, depending on whether or not they originate from a reconstructed $\Dstarp\to\Dp\piz$ decay. The tagged sample features a better signal-to-background ratio, while the null tag has larger signal yield. The \CP asymmetry is measured independently in each sample and the results are later combined. To avoid potential bias, the measured values of $A^{\pip\piz}$ remained undisclosed until the entire analysis procedure was finalized and all uncertainties were determined.

The Belle~II detector~\cite{b2tech,Kou:2018nap} operates at the SuperKEKB asymmetric-energy $\epem$ collider~\cite{Akai:2018mbz}. It has a cylindrical geometry and consists of a silicon vertex detector comprising two inner layers of pixel detectors and four outer layers of double-sided strip detectors, a 56-layer central drift chamber, a time-of-propagation detector, an aerogel ring-imaging Cherenkov detector, and an electromagnetic calorimeter made of CsI(Tl) crystals, all located inside a 1.5\,T superconducting solenoid. A flux return outside the solenoid is instrumented with resistive-plate chambers and plastic scintillator modules to detect muons and \KL mesons. For the data used in this measurement, only part of the second layer of the pixel detector, covering 15\% of the azimuthal angle, was installed. The $z$ axis of the laboratory frame is defined as the central axis of the solenoid, with its positive direction determined by the direction of the electron beam.

We use simulated event samples to identify sources of background, optimize selection criteria, determine fit models, and validate the analysis procedure. We generate $\epem\to\Upsilon(4S)$ events and simulate particle decays with \textsc{EvtGen}~\cite{Lange:2001uf} and \textsc{Pythia8}~\cite{Sjostrand:2014zea}; we generate continuum $\epem\to\qqbar$, where $q$ is a $u$, $d$, $c$, or $s$ quark, with \textsc{KKMC}~\cite{Jadach:1999vf} and \textsc{Pythia8}; we simulate final-state radiation with \textsc{Photos}~\cite{Barberio:1990ms,Barberio:1993qi}; and we simulate detector response using \textsc{Geant4}~\cite{Agostinelli:2002hh}. Beam backgrounds are taken into account by overlaying random trigger data.

Events are selected by a trigger based on either the total energy deposited in the calorimeter or the number of charged-particle tracks reconstructed in the central drift chamber. The efficiency of the trigger is close to 100\% for both signal and control mode decays. 
The off-line event reconstruction~\cite{Kuhr:2018lps,basf2-zenodo} starts by selecting events that are inconsistent with Bhabha scattering, and by requiring at least three charged particles that originate from the \epem interaction region, meaning that they have longitudinal and transverse distances of closest approach to the \epem interaction point (impact parameters) smaller than 3\cm and 1\cm, respectively, and have transverse momenta greater than 200\mevc.

Charged pion candidates must originate from the \epem interaction region and have tracks with hits in the central drift chamber, transverse momenta larger than 0.1\gevc, and momenta in the \epem center-of-mass system (c.m.s.)\ larger than 0.8\gevc. Charged particles are identified as pions with an efficiency of 98\%, and a kaon-to-pion misidentification rate of 27\%, using requirements on the output of a neural network that combines kinematic information, and particle-identification information from each subdetector~\cite{PID}. We reconstruct photon candidates from localized energy deposits (clusters) from at least two calorimeter crystals. The clusters should have polar angles within the acceptance of the drift chamber ($17<\theta<150^\circ$) to ensure that they are not matched to tracks. Clusters originating from beam-background particles, split-offs of hadronic showers, and track-cluster matching failures are suppressed using two multivariate discriminators, based on the time difference between the collision and reconstructed cluster, cluster-shape information~\cite{Longo:2020zqt}, the distance between the cluster and the nearest track, and pulse-shape discrimination~\cite{Belle-II:2024qod}. Pairs of photon candidates are combined to form neutral pion candidates. Neutral pions from the \Dp decay are referred to as ``hard'' pions, to distinguish them from the lower-momentum ``soft'' pions originating from the \Dstarp decay. Photons used to form hard-pion candidates must have energies greater than 80, 30, or 60\mev if reconstructed in the forward ($12.4<\theta<31.4^\circ$), barrel ($32.2<\theta<128.7^\circ$), or backward ($130.7<\theta<155.7^\circ$) regions of the calorimeter. Hard neutral pions must have a diphoton mass in the range $[120,145]\mevcc$ (the typical diphoton mass resolution is 7\mevcc) and c.m.s.\ momentum larger than 0.9\gevc. Photons used to form soft-pion candidates must have energies greater than 25\mev if reconstructed in the forward or barrel region of the calorimeter, and greater than 40\mev if reconstructed in the backward region. Soft neutral pions must have a diphoton mass in the range $[105,150]\mevcc$ and c.m.s.\ momentum larger than 0.1\gevc. Both hard and soft neutral pions are subject to a kinematic fit that constrains the diphoton mass to the known \piz mass~\cite{pdg} . Only candidates with successful fits and, for soft neutral pions, having \chisq probabilities larger than 0.01 are retained for subsequent analysis.

Candidate $\Dp\to\pip\piz$ decays are reconstructed from combinations of charged pions and hard neutral pions with invariant masses $m(\pip\piz)$ in the range $[1.6,2.3]\gevcc$. They are subject to a kinematic-vertex fit that constrains the \Dp production point to the measured position of the beam-interaction region~\cite{Krohn:2019dlq}. Only candidates with successful fits and having \chisq probabilities larger than 0.01 are retained. Candidate $\Dstarp\to\Dp\piz$ decays are reconstructed from combinations of \Dp and soft neutral-pion candidates. The difference between the masses of the \Dstarp and \Dp candidates, $\Delta m$, is required to be between 138 and 143\mevcc (the typical $\Delta m$ resolution is 2\mevcc). If more than one \Dstarp candidate is present for the same \Dp candidate, only the one with mass closest to the known \Dstarp mass is considered. To suppress events where the \Dp candidate comes from the decay of a \B meson, which may be affected by \CP violation in the \B decay, the c.m.s.\ momenta of the \Dp candidate in the null-tag sample and the \Dstarp candidate in the tagged sample are required to exceed 2.65 and 2.5\gevc, respectively.

An artificial neural network based on a multilayer perceptron is trained to suppress combinatorial background~\cite{Hocker:2007ht,TMVA2007} . The neural network is trained and tested on independent samples of simulated decays to prevent overtraining. The training is performed for null-tag candidates with $m(\pip\piz)$ in the range $[1.7,2.0]\gevcc$, to exclude background from partially reconstructed charm decays, and uses the following input variables: the asymmetry between the c.m.s.\ momenta of the \Dp final-state particles, their scalar sum, the logarithm of the charged pion transverse impact parameter, the charged pion longitudinal impact parameter, and the product between the reconstructed \D charge and the output of a charm-flavor tagger based on the rest of the $\epem\to\ccbar$ event~\cite{Belle-II:2023vra}. The input variables are chosen for their ability to distinguish between signal and background and for their similarity between signal and control modes. (For the control mode, the \piz variables are replaced with the corresponding \KS variables.) The transverse impact parameter provides the best discrimination, with signal decays having significantly more displaced charged pions compared to background because of the relatively long \Dp lifetime. This parameter also provides discrimination against background from \Dsp decays.

In the tagged sample, a requirement on the neural network suppresses the background in the $m(\pip\piz)$ range $[1.7,2.0]\gevcc$ by 78\%, while retaining 81\% of the signal. In the null-tag case, a tighter requirement on the neural network response, corresponding to 50\% signal efficiency for a background rejection of 98\%, is used. Backgrounds from $\Dsp\to\Kp\piz$ and $\Dsp\to\Kp\KS(\to\piz\piz)$ decays, where the charged kaon is misidentified as a pion and one $\piz$ from the \KS decay is not reconstructed, are suppressed by tightening the particle-identification requirement on the charged pions from null-tag \Dp candidates. The requirement has a pion-identification efficiency of 76.4\% for a kaon-to-pion misidentification rate of about 2.8\%. The remaining \Dsp contamination, about $0.4\%$, is small compared to the uncertainty in the signal yield, and is neglected.

The requirements on c.m.s.\ momenta, photon multivariate discriminators, kinematic-(vertex-)fit probabilities, and neural-network response are decided by maximizing the figure of merit $S/\sqrt{S+B}$, where $S$ and $B$ stand for the signal and background yields, respectively, in the $m(\pip\piz)$ range $[1.83,1.89]\gevcc$. After selection, about 1.5\% of events contain more than one \Dp candidate in both the tagged and null-tag samples. When this happens, only the candidate including the \piz candidate with the largest kinematic-fit \chisq probability is kept.

Control mode $\Dp\to\pip\KS$ decays are formed by combining charged-pion and \KS candidates. The \KS candidates are reconstructed from combinations of oppositely charged particles, which are assumed to be pions and are constrained to originate from a common vertex. The dipion mass is required to be in the range $[0.45,0.55]\gevcc$. The \KS flight length divided by its uncertainty must be larger than 10.0. Its c.m.s.\ momentum must be larger than 0.9\gevc. The $\Dp\to\pip\KS$ candidates are subject to the same kinematic-vertex fit and selected using the same requirements as for signal decays. They are similarly split into tagged and null-tag decays by combining with soft neutral pions. In the null-tag sample, dedicated vetoes remove background from $\Dstarp\to\Dz(\to\Km\pip)\pip$ and $\Lc\to\Lz(\to p\pim)\pip$, which occur through misidentification of kaons and protons as pions and are negligible in the tagged sample. The requirements on the particle-identification and background-suppression neural network outputs, both inherited from the selection of the signal mode, are sufficient to reduce to a negligible level the contamination of $\Dsp\to\Kp\KS$ decays, where the kaon is misidentified as a pion. 

\begin{figure*}[t]
\centering
\begin{overpic}[width=\textwidth]{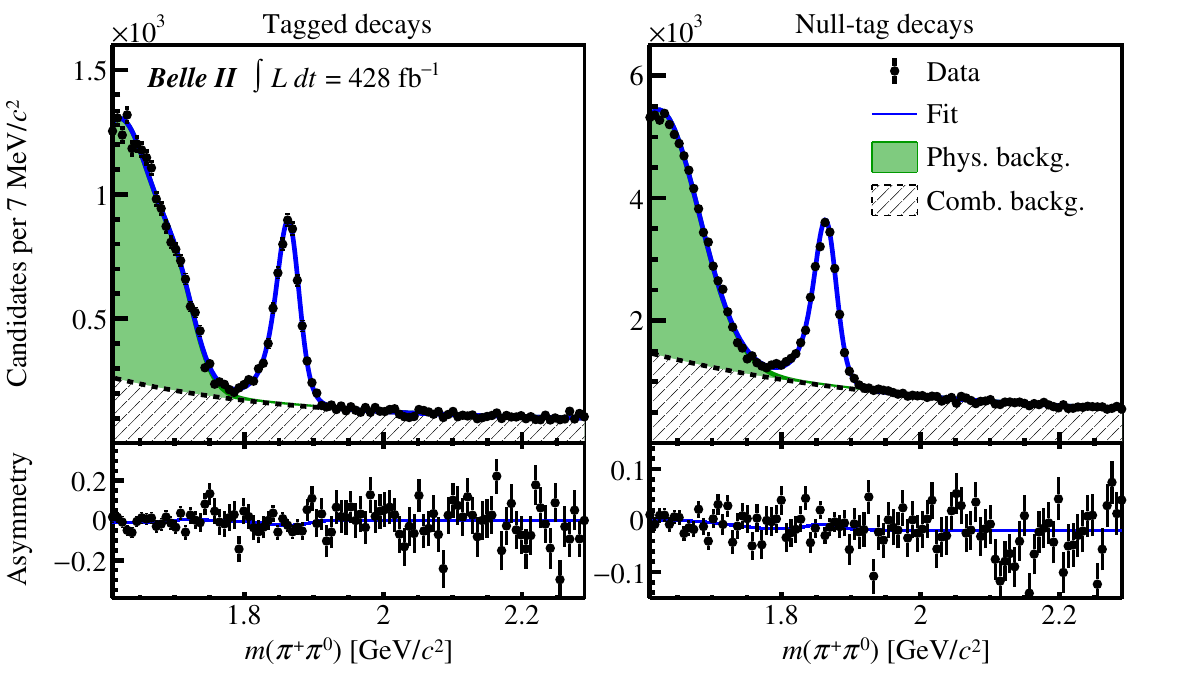}
\end{overpic}
\caption{Distributions of $m(\pip\piz)$ for (left) tagged and (right) null-tag $\Dp\to\pip\piz$ candidates, with fit projections overlaid. The bottom panels show the asymmetry as a function of mass, with fit projections overlaid. \label{fig:fit-signal-data}}
\end{figure*}

The raw asymmetries are determined from unbinned maximum-likelihood fits to the $m(\pip\piz)$ and $m(\pip\KS)$ distributions of the selected $\Dp\to\pip\piz$ and $\Dp\to\pip\KS$ candidates, split according to the \D meson charge. For $\Dp\to\pip\piz$ candidates, the fit considers three components: signal decays, physics background from misreconstructed charm decays, and combinatorial background. The signal probability density function (PDF) is modeled by the convolution of a Johnson's $S_{U}$ distribution~\cite{johnson} and a Gaussian distribution. The parameters of the Johnson's $S_{U}$ distribution are fixed to values obtained from simulation. The parameters of the Gaussian distribution are floated to account for possible data-simulation differences in peak position and resolution. The physics background is mainly composed of $\Dz\to\pip\pim\piz$ decays where one of the charged pions is not reconstructed; $\Dp\to\pip\piz\piz$ decays with a missing neutral pion; semileptonic decays such as $\Dp\to\piz\mup\nu$, where the muon is misidentified as a pion and the neutrino is not reconstructed; and $\Dp\to\KS(\to\piz\piz)\pip$ decays, where one neutral pion from the \KS decay is not reconstructed. The physics background populates the $m(\pip\piz)$ region below $1.8\gevcc$ and is modeled using a Gaussian function in the null-tag sample and two Gaussian functions in the tagged sample. The combinatorial background arises from accidental combinations of charged and neutral pion candidates. It has a smoothly falling distribution in $m(\pip\piz)$, which is modeled using the sum of an exponential PDF and a uniform distribution. All background parameters are floated in the fit. The other fit parameters are the yields and asymmetries of each component. The same models are used for \Dp and \Dm decays. The $m(\pip\piz)$ distributions of the $\Dp\to\pip\piz$ candidates are shown in \cref{fig:fit-signal-data}, with fit projections overlaid. The fit describes the data fairly well. The signal yields are determined to be $5\,130\pm110$ and $18\,510\pm240$ in the tagged and null-tag samples, respectively. The raw asymmetries are $(-2.9 \pm 1.8)\%$ and $(-0.4 \pm 1.0)\%$, respectively.  The uncertainties are statistical only.

\begin{figure*}[t]
\centering
\begin{overpic}[width=\textwidth]{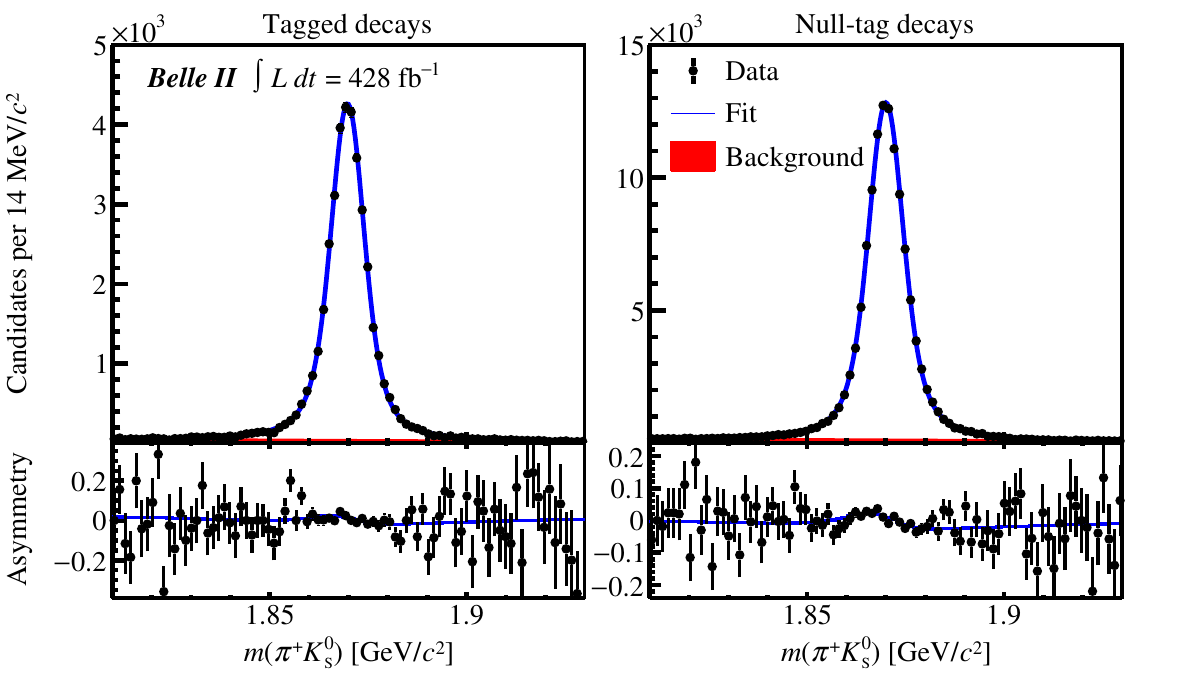}
\end{overpic}
\caption{Distributions of $m(\pip\KS)$ for (left) tagged and (right) null-tag $\Dp\to\pip\KS$ candidates, with fit projections overlaid. The bottom panels show the asymmetry as a function of mass, with fit projections overlaid. \label{fig:fit-control-data}}
\end{figure*}

The fit to the $m(\pip\KS)$ distributions of the control sample considers the $\Dp\to\pip\KS$ component, modeled as a Johnson's $S_{U}$ distribution convolved with a Gaussian function, and a background component, modeled by an exponential distribution. The width and the mean of the Johnson's $S_U$ distribution are allowed to differ between \Dp and \Dm candidates, to account for small differences in momentum scale and resolution of positively and negatively charged particles. (Differences in \Dp and \Dm shapes are diluted in the signal mode because the mass scale and resolution are dominated by the energy scale and resolution of the neutral pion.) All parameters are floated in the fit. The fit describes the data well, as shown in \cref{fig:fit-control-data}. The $\Dp\to\pip\KS$ yields are determined to be $39\,630\pm300$ and $123\,560 \pm 500$ in the tagged and null-tag samples, respectively. The raw asymmetries are $(0.54 \pm 0.53)\%$ and $(0.33 \pm 0.30)\%$, respectively. The uncertainties are statistical only.

The $A^{\Kzb}$ contributions to the $\Dp\to\pip\KS$ raw asymmetries are computed following Ref.~\cite{LHCb:2014kcb}. The computation uses the \KS candidate flight lengths and directions, the well-known mixing and \CP-violation parameters of the \Kz-\Kzb system, the well-known interaction cross sections, and the detector material density. We estimate $A^{\Kzb}$ to be $(-0.422\pm0.007)\%$ and $(-0.418\pm0.007)\%$ for tagged and null-tag samples, respectively. The uncertainties are mainly systematic and due to the detector material density, which is known with a relative $5\%$ uncertainty.

Using \cref{eq:Acp}, we compute the values of $\Acp(\Dp\to\pip\piz)$ in the tagged and null-tag samples to be $(\resValueTagg\pm\resStatTagg)\%$ and $(\resValueNull\pm\resStatNull)\%$, respectively, where the uncertainties are statistical only. The results are consistent with each other. The analysis is validated using sets of pseudoexperiments generated by sampling from the fit PDFs and using fully simulated events, which confirm that we estimate $\Acp(\Dp\to\pip\piz)$, and its uncertainty, without bias. Performing the measurement in independent subsets of the data, selected according to varying data-taking conditions, and varying \Dp momentum, polar, and azimuthal angle ranges, returns consistent $\Acp(\Dp\to\pip\piz)$ results.

Three sources of systematic uncertainties are considered: modeling of the mass distributions in the fits (for both signal and control modes),  neglected differences between the kinematic distributions of the signal and the control modes, and uncertainty in the neutral kaon asymmetry. 

To estimate the systematic uncertainty due to the fit models, we repeat the fits to the data using alternative models that give an equally good description of the data. In the $\Dp\to\pip\piz$ fit, the signal PDF is modified to a Johnson's $S_{U}$ distribution (\ie, without convolving with a Gaussian function as in the default model) with floated parameters; the combinatorial background model is changed to a single exponential distribution; and the physics background model is modified by adding a Gaussian distribution in the null-tag sample, and by using a Johnson's $S_{U}$ distribution in the tagged sample. For each variation, the shifts in the measured values of $A^{\pip\piz}$ with respect to the default results are computed. The sums in quadrature of these shifts, 0.119\% for the tagged sample and 0.044\% for the null-tag sample, are assigned as systematic uncertainties due to the $\Dp\to\pip\piz$ fit model. In the control-mode fit, we replace the $\Dp\to\pip\KS$ PDF with a sum of a Johnson's $S_{U}$ distribution and a Gaussian function, and the background PDF with a straight line or with the sum of an exponential and a constant. In the latter case, the fit range is also extended to 2.3\gevcc, to better constrain the background. For the null-tag case, extending the fit range requires the inclusion of a fit component for $\Dsp\to\pip\KS$ decays, which is modeled using the same shape as for $\Dp\to\pip\KS$ decays except for an overall mass shift. The sums in quadrature of the shifts in $A^{\pip\KS}$ with respect to the default results, $0.122\%$ for the tagged sample and $0.048\%$ for the null-tag sample, are assigned as systematic uncertainties. The default fit models assume that most shape parameters are charge independent. To verify this assumption, we refit to the data by replacing individual shape parameters with charge-dependent ones. In all cases, the parameter asymmetries are consistent with zero and statistically insignificant shifts are observed in the measured raw asymmetries.

The subtraction of raw asymmetries between signal and control decays precisely cancels the contributions from production and detection asymmetries only if signal and control decays have similar kinematic distributions. In particular, the \Dp polar angle distributions in the c.m.s.\ must agree between the signal and control modes to cancel the production asymmetry; and the kinematic distributions of the charged pion must agree to cancel the charged-pion detection asymmetry. The control samples are weighted to correct for observed small differences in these kinematic distributions. The weighting reduces the effective sizes of the tagged and null-tag $\Dp\to\pip\KS$ samples by 1.2\% and 2.0\%, respectively. The weighted control-sample data are then fit and the absolute shifts in the measured values of $A^{\pip\KS}$, $0.096\%$ in the tagged sample and $0.053\%$ in the null-tag sample, are assigned as systematic uncertainties.

The uncertainties in $A^{\Kzb}$, $0.007\%$ for both samples, are assigned as systematic uncertainties in $\Acp(\Dp\to\pip\piz)$. The total systematic uncertainties, $0.196\%$ for the tagged sample and $0.084\%$ for the null-tag sample, are the sums in quadrature of the individual components.

In conclusion, we measure the \CP asymmetry in $\Dp\to\pip\piz$ decays using a sample of $\epem\to\ccbar$ data collected by Belle II, which has an integrated luminosity of 428\invfb. The sample is split according to whether the \Dp meson arises from a reconstructed $\Dstarp\to\Dp\piz$ decay or not. 
The \CP asymmetries in the two samples are measured to be $(\resValueTagg\pm\resStatTagg\pm\resSystTagg)\%$ and $(\resValueNull\pm\resStatNull\pm\resSystNull)\%$, respectively, where the first uncertainties are statistical and the second systematic. They agree with each other and are combined to obtain
\begin{equation}
\Acp(D^+ \to \pi^+ \pi^0) = (\resValueComb\pm\resStatComb\pm\resSystComb)\%\,.
\end{equation}
In the combination, we assumed uncorrelated systematic uncertainties due to the fit models and fully correlated uncertainties due to the kinematic weighting and neutral kaon asymmetry. The result agrees with \CP symmetry and with previous measurements~\cite{CLEO:2009fiz,Belle:2017tho,LHCb:2021rou}. The 30\% improved precision compared to Belle's result, based on twice as much integrated luminosity~\cite{Belle:2017tho}, is due to the substantially better sample purity achieved through an improved event selection, which exploits Belle~II's superior performance in the reconstruction of neutral pions and displaced charged particles.  The result is the most precise measurement to date.
 
\begin{acknowledgments}
%
%
This work, based on data collected using the Belle II detector, which was built and commissioned prior to March 2019,
was supported by
Higher Education and Science Committee of the Republic of Armenia Grant No.~23LCG-1C011;
Australian Research Council and Research Grants
No.~DP200101792, %
No.~DP210101900, %
No.~DP210102831, %
No.~DE220100462, %
No.~LE210100098, %
and
No.~LE230100085; %
Austrian Federal Ministry of Education, Science and Research,
Austrian Science Fund (FWF) Grants
DOI:~10.55776/P34529,
DOI:~10.55776/J4731,
DOI:~10.55776/J4625,
DOI:~10.55776/M3153,
and
DOI:~10.55776/PAT1836324,
and
Horizon 2020 ERC Starting Grant No.~947006 ``InterLeptons'';
Natural Sciences and Engineering Research Council of Canada, Compute Canada and CANARIE;
National Key R\&D Program of China under Contract No.~2024YFA1610503,
and
No.~2024YFA1610504
National Natural Science Foundation of China and Research Grants
No.~11575017,
No.~11761141009,
No.~11705209,
No.~11975076,
No.~12135005,
No.~12150004,
No.~12161141008,
No.~12475093,
and
No.~12175041,
and Shandong Provincial Natural Science Foundation Project~ZR2022JQ02;
the Czech Science Foundation Grant No. 22-18469S,  Regional funds of EU/MEYS: OPJAK
FORTE CZ.02.01.01/00/22\_008/0004632 
and
Charles University Grant Agency project No. 246122;
European Research Council, Seventh Framework PIEF-GA-2013-622527,
Horizon 2020 ERC-Advanced Grants No.~267104 and No.~884719,
Horizon 2020 ERC-Consolidator Grant No.~819127,
Horizon 2020 Marie Sklodowska-Curie Grant Agreement No.~700525 ``NIOBE''
and
No.~101026516,
and
Horizon 2020 Marie Sklodowska-Curie RISE project JENNIFER2 Grant Agreement No.~822070 (European grants);
L'Institut National de Physique Nucl\'{e}aire et de Physique des Particules (IN2P3) du CNRS
and
L'Agence Nationale de la Recherche (ANR) under Grant No.~ANR-21-CE31-0009 (France);
BMFTR, DFG, HGF, MPG, and AvH Foundation (Germany);
Department of Atomic Energy under Project Identification No.~RTI 4002,
Department of Science and Technology,
and
UPES SEED funding programs
No.~UPES/R\&D-SEED-INFRA/17052023/01 and
No.~UPES/R\&D-SOE/20062022/06 (India);
Israel Science Foundation Grant No.~2476/17,
U.S.-Israel Binational Science Foundation Grant No.~2016113, and
Israel Ministry of Science Grant No.~3-16543;
Istituto Nazionale di Fisica Nucleare and the Research Grants BELLE2,
and
the ICSC – Centro Nazionale di Ricerca in High Performance Computing, Big Data and Quantum Computing, funded by European Union – NextGenerationEU;
Japan Society for the Promotion of Science, Grant-in-Aid for Scientific Research Grants
No.~16H03968,
No.~16H03993,
No.~16H06492,
No.~16K05323,
No.~17H01133,
No.~17H05405,
No.~18K03621,
No.~18H03710,
No.~18H05226,
No.~19H00682, %
No.~20H05850,
No.~20H05858,
No.~22H00144,
No.~22K14056,
No.~22K21347,
No.~23H05433,
No.~26220706,
and
No.~26400255,
and
the Ministry of Education, Culture, Sports, Science, and Technology (MEXT) of Japan;  
National Research Foundation (NRF) of Korea Grants
No.~2021R1-F1A-1064008, 
No.~2022R1-A2C-1003993,
No.~2022R1-A2C-1092335,
No.~RS-2016-NR017151,
No.~RS-2018-NR031074,
No.~RS-2021-NR060129,
No.~RS-2023-00208693,
No.~RS-2024-00354342
and
No.~RS-2025-02219521,
Radiation Science Research Institute,
Foreign Large-Size Research Facility Application Supporting project,
the Global Science Experimental Data Hub Center, the Korea Institute of Science and
Technology Information (K25L2M2C3 ) 
and
KREONET/GLORIAD;
Universiti Malaya RU grant, Akademi Sains Malaysia, and Ministry of Education Malaysia;
Frontiers of Science Program Contracts
No.~FOINS-296,
No.~CB-221329,
No.~CB-236394,
No.~CB-254409,
and
No.~CB-180023, and SEP-CINVESTAV Research Grant No.~237 (Mexico);
the Polish Ministry of Science and Higher Education and the National Science Center;
the Ministry of Science and Higher Education of the Russian Federation
and
the HSE University Basic Research Program, Moscow;
University of Tabuk Research Grants
No.~S-0256-1438 and No.~S-0280-1439 (Saudi Arabia), and
Researchers Supporting Project number (RSPD2025R873), King Saud University, Riyadh,
Saudi Arabia;
Slovenian Research Agency and Research Grants
No.~J1-50010
and
No.~P1-0135;
Ikerbasque, Basque Foundation for Science,
State Agency for Research of the Spanish Ministry of Science and Innovation through Grant No. PID2022-136510NB-C33, Spain,
Agencia Estatal de Investigacion, Spain
Grant No.~RYC2020-029875-I
and
Generalitat Valenciana, Spain
Grant No.~CIDEGENT/2018/020;
The Knut and Alice Wallenberg Foundation (Sweden), Contracts No.~2021.0174 and No.~2021.0299;
National Science and Technology Council,
and
Ministry of Education (Taiwan);
Thailand Center of Excellence in Physics;
TUBITAK ULAKBIM (Turkey);
National Research Foundation of Ukraine, Project No.~2020.02/0257,
and
Ministry of Education and Science of Ukraine;
the U.S. National Science Foundation and Research Grants
No.~PHY-1913789 %
and
No.~PHY-2111604, %
and the U.S. Department of Energy and Research Awards
No.~DE-AC06-76RLO1830, %
No.~DE-SC0007983, %
No.~DE-SC0009824, %
No.~DE-SC0009973, %
No.~DE-SC0010007, %
No.~DE-SC0010073, %
No.~DE-SC0010118, %
No.~DE-SC0010504, %
No.~DE-SC0011784, %
No.~DE-SC0012704, %
No.~DE-SC0019230, %
No.~DE-SC0021274, %
No.~DE-SC0021616, %
No.~DE-SC0022350, %
No.~DE-SC0023470; %
and
the Vietnam Academy of Science and Technology (VAST) under Grants
No.~NVCC.05.12/22-23
and
No.~DL0000.02/24-25.

These acknowledgements are not to be interpreted as an endorsement of any statement made
by any of our institutes, funding agencies, governments, or their representatives.

We thank the SuperKEKB team for delivering high-luminosity collisions;
the KEK cryogenics group for the efficient operation of the detector solenoid magnet and IBBelle on site;
the KEK Computer Research Center for on-site computing support; the NII for SINET6 network support;
and the raw-data centers hosted by BNL, DESY, GridKa, IN2P3, INFN, 
and the University of Victoria.
 \end{acknowledgments}

\bibliographystyle{belle2}

\begin{thebibliography}{10}

\bibitem{Cabibbo:1963yz}
N.~Cabibbo, \ifthenelse{\boolean{articletitles}}{\emph{{Unitary symmetry and
  leptonic decays}},
  }{}\href{https://doi.org/10.1103/PhysRevLett.10.531}{Phys.\ Rev.\ Lett.\
  \textbf{10} (1963) 531}.

\bibitem{Kobayashi:1973fv}
M.~Kobayashi and T.~Maskawa,
  \ifthenelse{\boolean{articletitles}}{\emph{{\CP-violation in the
  renormalizable theory of weak interaction}},
  }{}\href{https://doi.org/10.1143/PTP.49.652}{Prog.\ Theor.\ Phys.\
  \textbf{49} (1973) 652}.

\bibitem{Christenson:1964fg}
J.~H. Christenson, J.~W. Cronin, V.~L. Fitch, and R.~Turlay,
  \ifthenelse{\boolean{articletitles}}{\emph{{Evidence for the $2\pi$ decay of
  the $K_2^0$ meson}},
  }{}\href{https://doi.org/10.1103/PhysRevLett.13.138}{Phys.\ Rev.\ Lett.\
  \textbf{13} (1964) 138}.

\bibitem{KTeV:1999kad}
KTeV collaboration, A.~Alavi-Harati {\em et~al.},
  \ifthenelse{\boolean{articletitles}}{\emph{{Observation of Direct \CP
  Violation in $K_{S,L} \to \pi \pi$ Decays}},
  }{}\href{https://doi.org/10.1103/PhysRevLett.83.22}{Phys.\ Rev.\ Lett.\
  \textbf{83} (1999) 22},
  \href{http://arxiv.org/abs/hep-ex/9905060}{{\normalfont\ttfamily
  arXiv:hep-ex/9905060}}.

\bibitem{NA48:1999szy}
NA48 collaboration, V.~Fanti {\em et~al.},
  \ifthenelse{\boolean{articletitles}}{\emph{{A new measurement of direct \CP
  violation in two pion decays of the neutral kaon}},
  }{}\href{https://doi.org/10.1016/S0370-2693(99)01030-8}{Phys.\ Lett.\ B
  \textbf{465} (1999) 335},
  \href{http://arxiv.org/abs/hep-ex/9909022}{{\normalfont\ttfamily
  arXiv:hep-ex/9909022}}.

\bibitem{BaBar:2002kla}
BaBar collaboration, B.~Aubert {\em et~al.},
  \ifthenelse{\boolean{articletitles}}{\emph{{Measurement of the \CP-violating
  asymmetry amplitude $\sin 2\beta$}},
  }{}\href{https://doi.org/10.1103/PhysRevLett.89.201802}{Phys.\ Rev.\ Lett.\
  \textbf{89} (2002) 201802},
  \href{http://arxiv.org/abs/hep-ex/0207042}{{\normalfont\ttfamily
  arXiv:hep-ex/0207042}}.

\bibitem{Belle:2002ghj}
Belle collaboration, K.~Abe {\em et~al.},
  \ifthenelse{\boolean{articletitles}}{\emph{{An Improved measurement of mixing
  induced \CP violation in the neutral \B meson system}},
  }{}\href{https://doi.org/10.1103/PhysRevD.66.071102}{Phys.\ Rev.\ D
  \textbf{66} (2002) 071102},
  \href{http://arxiv.org/abs/hep-ex/0208025}{{\normalfont\ttfamily
  arXiv:hep-ex/0208025}}.

\bibitem{Aubert:2004qm}
BaBar collaboration, B.~Aubert {\em et~al.},
  \ifthenelse{\boolean{articletitles}}{\emph{{Direct \CP violating asymmetry in
  \mbox{$\Bz\to K^+ \pi^-$} decays}},
  }{}\href{https://doi.org/10.1103/PhysRevLett.93.131801}{Phys.\ Rev.\ Lett.\
  \textbf{93} (2004) 131801},
  \href{http://arxiv.org/abs/hep-ex/0407057}{{\normalfont\ttfamily
  arXiv:hep-ex/0407057}}.

\bibitem{Chao:2004mn}
Belle collaboration, Y.~Chao {\em et~al.},
  \ifthenelse{\boolean{articletitles}}{\emph{{Evidence for direct \CP violation
  in $B^0\to \Kp \pim$ decays}},
  }{}\href{https://doi.org/10.1103/PhysRevLett.93.191802}{Phys.\ Rev.\ Lett.\
  \textbf{93} (2004) 191802},
  \href{http://arxiv.org/abs/hep-ex/0408100}{{\normalfont\ttfamily
  arXiv:hep-ex/0408100}}.

\bibitem{LHCb-PAPER-2012-001}
LHCb collaboration, R.~Aaij {\em et~al.},
  \ifthenelse{\boolean{articletitles}}{\emph{{Observation of \CP violation in
  $\Bpm\to \D\Kpm$ decays}},
  }{}\href{https://doi.org/10.1016/j.physletb.2012.04.060}{Phys.\ Lett.\ B
  \textbf{712} (2012) 203}, Erratum
  \href{https://doi.org/10.1016/j.physletb.2012.05.060}{ibid.\   \textbf{B713}
  (2012) 351}, \href{http://arxiv.org/abs/1203.3662}{{\normalfont\ttfamily
  arXiv:1203.3662}}.

\bibitem{LHCb-PAPER-2013-018}
LHCb collaboration, R.~Aaij {\em et~al.},
  \ifthenelse{\boolean{articletitles}}{\emph{{First observation of \CP
  violation in the decays of $\Bs$ mesons}},
  }{}\href{https://doi.org/10.1103/PhysRevLett.110.221601}{Phys.\ Rev.\ Lett.\
  \textbf{110} (2013) 221601},
  \href{http://arxiv.org/abs/1304.6173}{{\normalfont\ttfamily
  arXiv:1304.6173}}.

\bibitem{Glashow:1970gm}
S.~L. Glashow, J.~Iliopoulos, and L.~Maiani,
  \ifthenelse{\boolean{articletitles}}{\emph{{Weak Interactions with
  Lepton-Hadron Symmetry}},
  }{}\href{https://doi.org/10.1103/PhysRevD.2.1285}{Phys.\ Rev.\ D \textbf{2}
  (1970) 1285}.

\bibitem{Golden:1989qx}
M.~Golden and B.~Grinstein,
  \ifthenelse{\boolean{articletitles}}{\emph{{Enhanced \CP violations in
  hadronic charm decays}},
  }{}\href{https://doi.org/10.1016/0370-2693(89)90353-5}{Phys.\ Lett.\ B
  \textbf{222} (1989) 501}.

\bibitem{Buccella:1994nf}
F.~Buccella {\em et~al.},
  \ifthenelse{\boolean{articletitles}}{\emph{{Nonleptonic weak decays of
  charmed mesons}}, }{}\href{https://doi.org/10.1103/PhysRevD.51.3478}{Phys.\
  Rev.\ D \textbf{51} (1995) 3478},
  \href{http://arxiv.org/abs/hep-ph/9411286}{{\normalfont\ttfamily
  arXiv:hep-ph/9411286}}.

\bibitem{Bianco:2003vb}
S.~Bianco, F.~L. Fabbri, D.~Benson, and I.~Bigi,
  \ifthenelse{\boolean{articletitles}}{\emph{{A Cicerone for the physics of
  charm}}, }{}\href{https://doi.org/10.1393/ncr/i2003-10003-1}{Riv.\ Nuovo
  Cim.\  \textbf{26N7} (2003) 1},
  \href{http://arxiv.org/abs/hep-ex/0309021}{{\normalfont\ttfamily
  arXiv:hep-ex/0309021}}.

\bibitem{Grossman:2006jg}
Y.~Grossman, A.~L. Kagan, and Y.~Nir,
  \ifthenelse{\boolean{articletitles}}{\emph{{New physics and \CP violation in
  singly Cabibbo suppressed $D$ decays}},
  }{}\href{https://doi.org/10.1103/PhysRevD.75.036008}{Phys.\ Rev.\ D
  \textbf{75} (2007) 036008},
  \href{http://arxiv.org/abs/hep-ph/0609178}{{\normalfont\ttfamily
  arXiv:hep-ph/0609178}}.

\bibitem{Artuso:2008vf}
M.~Artuso, B.~Meadows, and A.~A. Petrov,
  \ifthenelse{\boolean{articletitles}}{\emph{{Charm meson decays}},
  }{}\href{https://doi.org/10.1146/annurev.nucl.58.110707.171131}{Ann.\ Rev.\
  Nucl.\ Part.\ Sci.\  \textbf{58} (2008) 249},
  \href{http://arxiv.org/abs/0802.2934}{{\normalfont\ttfamily
  arXiv:0802.2934}}.

\bibitem{Aaij:2019kcg}
LHCb collaboration, R.~Aaij {\em et~al.},
  \ifthenelse{\boolean{articletitles}}{\emph{{Observation of \CP violation in
  charm decays}},
  }{}\href{https://doi.org/10.1103/PhysRevLett.122.211803}{Phys.\ Rev.\ Lett.\
  \textbf{122} (2019) 211803},
  \href{http://arxiv.org/abs/1903.08726}{{\normalfont\ttfamily
  arXiv:1903.08726}}.

\bibitem{LHCb:2022lry}
LHCb collaboration, R.~Aaij {\em et~al.},
  \ifthenelse{\boolean{articletitles}}{\emph{{Measurement of the
  time-integrated \CP asymmetry in $\Dz\to\Kp\Km$ decays}},
  }{}\href{https://doi.org/10.1103/PhysRevLett.131.091802}{Phys.\ Rev.\ Lett.\
  \textbf{131} (2023) 091802},
  \href{http://arxiv.org/abs/2209.03179}{{\normalfont\ttfamily
  arXiv:2209.03179}}.

\bibitem{Chala:2019fdb}
M.~Chala, A.~Lenz, A.~V. Rusov, and J.~Scholtz,
  \ifthenelse{\boolean{articletitles}}{\emph{{$\Delta A_{\CP}$ within the
  Standard Model and beyond}},
  }{}\href{https://doi.org/10.1007/JHEP07(2019)161}{JHEP \textbf{07} (2019)
  161}, \href{http://arxiv.org/abs/1903.10490}{{\normalfont\ttfamily
  arXiv:1903.10490}}.

\bibitem{Dery:2019ysp}
A.~Dery and Y.~Nir, \ifthenelse{\boolean{articletitles}}{\emph{{Implications of
  the LHCb discovery of \CP violation in charm decays}},
  }{}\href{https://doi.org/10.1007/JHEP12(2019)104}{JHEP \textbf{12} (2019)
  104}, \href{http://arxiv.org/abs/1909.11242}{{\normalfont\ttfamily
  arXiv:1909.11242}}.

\bibitem{Calibbi:2019bay}
L.~Calibbi, T.~Li, Y.~Li, and B.~Zhu,
  \ifthenelse{\boolean{articletitles}}{\emph{{Simple model for large \CP
  violation in charm decays, $B$-physics anomalies, muon $g-2$ and dark
  matter}}, }{}\href{https://doi.org/10.1007/JHEP10(2020)070}{JHEP \textbf{10}
  (2020) 070}, \href{http://arxiv.org/abs/1912.02676}{{\normalfont\ttfamily
  arXiv:1912.02676}}.

\bibitem{Grossman:2019xcj}
Y.~Grossman and S.~Schacht, \ifthenelse{\boolean{articletitles}}{\emph{{The
  emergence of the $\Delta U=0$ rule in charm physics}},
  }{}\href{https://doi.org/10.1007/JHEP07(2019)020}{JHEP \textbf{07} (2019)
  020}, \href{http://arxiv.org/abs/1903.10952}{{\normalfont\ttfamily
  arXiv:1903.10952}}.

\bibitem{Cheng:2019ggx}
H.-Y. Cheng and C.-W. Chiang,
  \ifthenelse{\boolean{articletitles}}{\emph{{Revisiting \CP violation in $D\to
  P\!P$ and $V\!P$ decays}},
  }{}\href{https://doi.org/10.1103/PhysRevD.100.093002}{Phys.\ Rev.\ D
  \textbf{100} (2019) 093002},
  \href{http://arxiv.org/abs/1909.03063}{{\normalfont\ttfamily
  arXiv:1909.03063}}.

\bibitem{Buras:2021rdg}
A.~J. Buras, P.~Colangelo, F.~De~Fazio, and F.~Loparco,
  \ifthenelse{\boolean{articletitles}}{\emph{{The charm of 331}},
  }{}\href{https://doi.org/10.1007/JHEP10(2021)021}{JHEP \textbf{10} (2021)
  021}, \href{http://arxiv.org/abs/2107.10866}{{\normalfont\ttfamily
  arXiv:2107.10866}}.

\bibitem{Schacht:2021jaz}
S.~Schacht and A.~Soni, \ifthenelse{\boolean{articletitles}}{\emph{{Enhancement
  of charm \CP violation due to nearby resonances}},
  }{}\href{https://doi.org/10.1016/j.physletb.2021.136855}{Phys.\ Lett.\ B
  \textbf{825} (2022) 136855},
  \href{http://arxiv.org/abs/2110.07619}{{\normalfont\ttfamily
  arXiv:2110.07619}}.

\bibitem{Bediaga:2022sxw}
I.~Bediaga, T.~Frederico, and P.~C. Magalh\~aes,
  \ifthenelse{\boolean{articletitles}}{\emph{{Enhanced Charm \CP Asymmetries
  from Final State Interactions}},
  }{}\href{https://doi.org/10.1103/PhysRevLett.131.051802}{Phys.\ Rev.\ Lett.\
  \textbf{131} (2023) 051802},
  \href{http://arxiv.org/abs/2203.04056}{{\normalfont\ttfamily
  arXiv:2203.04056}}.

\bibitem{Bause:2022jes}
R.~Bause {\em et~al.}, \ifthenelse{\boolean{articletitles}}{\emph{{$U$-spin-\CP
  anomaly in charm}},
  }{}\href{https://doi.org/10.1103/PhysRevD.108.035005}{Phys.\ Rev.\ D
  \textbf{108} (2023) 035005},
  \href{http://arxiv.org/abs/2210.16330}{{\normalfont\ttfamily
  arXiv:2210.16330}}.

\bibitem{Schacht:2022kuj}
S.~Schacht, \ifthenelse{\boolean{articletitles}}{\emph{{A $U$-spin anomaly in
  charm \CP violation}}, }{}\href{https://doi.org/10.1007/JHEP03(2023)205}{JHEP
  \textbf{03} (2023) 205},
  \href{http://arxiv.org/abs/2207.08539}{{\normalfont\ttfamily
  arXiv:2207.08539}}.

\bibitem{Pich:2023kim}
A.~Pich, E.~Solomonidi, and L.~Vale~Silva,
  \ifthenelse{\boolean{articletitles}}{\emph{{Final-state interactions in the
  CP asymmetries of charm-meson two-body decays}},
  }{}\href{https://doi.org/10.1103/PhysRevD.108.036026}{Phys.\ Rev.\ D
  \textbf{108} (2023) 036026},
  \href{http://arxiv.org/abs/2305.11951}{{\normalfont\ttfamily
  arXiv:2305.11951}}.

\bibitem{Gavrilova:2023fzy}
M.~Gavrilova, Y.~Grossman, and S.~Schacht,
  \ifthenelse{\boolean{articletitles}}{\emph{{Determination of the $D\to\pi\pi$
  ratio of penguin over tree diagrams}},
  }{}\href{https://doi.org/10.1103/PhysRevD.109.033011}{Phys.\ Rev.\ D
  \textbf{109} (2024) 033011},
  \href{http://arxiv.org/abs/2312.10140}{{\normalfont\ttfamily
  arXiv:2312.10140}}.

\bibitem{Grossman:2012}
Y.~Grossman, A.~L. Kagan, and J.~Zupan,
  \ifthenelse{\boolean{articletitles}}{\emph{{Testing for new physics in singly
  Cabibbo suppressed \D decays}},
  }{}\href{https://doi.org/10.1103/PhysRevD.85.114036}{Phys.\ Rev.\ D
  \textbf{85} (2012) 114036},
  \href{http://arxiv.org/abs/1204.3557}{{\normalfont\ttfamily
  arXiv:1204.3557}}.

\bibitem{Hiller:2012xm}
G.~Hiller, M.~Jung, and S.~Schacht,
  \ifthenelse{\boolean{articletitles}}{\emph{{$SU(3)$-flavor anatomy of
  nonleptonic charm decays}},
  }{}\href{https://doi.org/10.1103/PhysRevD.87.014024}{Phys.\ Rev.\ D
  \textbf{87} (2013) 014024},
  \href{http://arxiv.org/abs/1211.3734}{{\normalfont\ttfamily
  arXiv:1211.3734}}.

\bibitem{Buccella:2019kpn}
F.~Buccella, A.~Paul, and P.~Santorelli,
  \ifthenelse{\boolean{articletitles}}{\emph{{$SU(3)_F$ breaking through final
  state interactions and \CP asymmetries in $D \to PP$ decays}},
  }{}\href{https://doi.org/10.1103/PhysRevD.99.113001}{Phys.\ Rev.\ D
  \textbf{99} (2019) 113001},
  \href{http://arxiv.org/abs/1902.05564}{{\normalfont\ttfamily
  arXiv:1902.05564}}.

\bibitem{Buccella:1992sg}
F.~Buccella {\em et~al.}, \ifthenelse{\boolean{articletitles}}{\emph{{\CP
  Violating asymmetries in charged \D meson decays}},
  }{}\href{https://doi.org/10.1016/0370-2693(93)90402-4}{Phys.\ Lett.\ B
  \textbf{302} (1993) 319},
  \href{http://arxiv.org/abs/hep-ph/9212253}{{\normalfont\ttfamily
  arXiv:hep-ph/9212253}}.

\bibitem{CLEO:2009fiz}
CLEO collaboration, H.~Mendez {\em et~al.},
  \ifthenelse{\boolean{articletitles}}{\emph{{Measurements of \D Meson Decays
  to Two Pseudoscalar Mesons}},
  }{}\href{https://doi.org/10.1103/PhysRevD.81.052013}{Phys.\ Rev.\ D
  \textbf{81} (2010) 052013},
  \href{http://arxiv.org/abs/0906.3198}{{\normalfont\ttfamily
  arXiv:0906.3198}}.

\bibitem{Belle:2017tho}
Belle collaboration, V.~Babu {\em et~al.},
  \ifthenelse{\boolean{articletitles}}{\emph{{Search for \CP violation in the
  $D^{+}\to\pi^{+}\pi^{0}$ decay at Belle}},
  }{}\href{https://doi.org/10.1103/PhysRevD.97.011101}{Phys.\ Rev.\ D
  \textbf{97} (2018) 011101},
  \href{http://arxiv.org/abs/1712.00619}{{\normalfont\ttfamily
  arXiv:1712.00619}}.

\bibitem{LHCb:2021rou}
LHCb collaboration, R.~Aaij {\em et~al.},
  \ifthenelse{\boolean{articletitles}}{\emph{{Search for \CP violation in $
  {D}_{(s)}^{+}\to {h}^{+}{\pi}^0 $ and $ {D}_{(s)}^{+}\to {h}^{+}\eta $
  decays}}, }{}\href{https://doi.org/10.1007/JHEP06(2021)019}{JHEP \textbf{06}
  (2021) 019}, \href{http://arxiv.org/abs/2103.11058}{{\normalfont\ttfamily
  arXiv:2103.11058}}.

\bibitem{lumi}
Belle~II collaboration, I.~Adachi {\em et~al.},
  \ifthenelse{\boolean{articletitles}}{\emph{{Measurement of the integrated
  luminosity of data samples collected during 2019-2022 by the Belle~II
  experiment}}, }{}\href{https://doi.org/10.1088/1674-1137/ad806c}{Chin.\
  Phys.\ C \textbf{49} (2025) 013001},
  \href{http://arxiv.org/abs/2407.00965}{{\normalfont\ttfamily
  arXiv:2407.00965}}.

\bibitem{Berends:1973fd}
F.~A. Berends, K.~J.~F. Gaemers, and R.~Gastmans,
  \ifthenelse{\boolean{articletitles}}{\emph{{$\alpha^3$ contribution to the
  angular asymmetry in $e^+e^-\to\mu^+\mu^-$}},
  }{}\href{https://doi.org/10.1016/0550-3213(73)90153-3}{Nucl.\ Phys.\ B
  \textbf{63} (1973) 381}.

\bibitem{Brown:1973ji}
R.~W. Brown, K.~O. Mikaelian, V.~K. Cung, and E.~A. Paschos,
  \ifthenelse{\boolean{articletitles}}{\emph{{Electromagnetic background in the
  search for neutral weak currents via $e^+e^-\to\mu^+\mu^-$}},
  }{}\href{https://doi.org/10.1016/0370-2693(73)90384-5}{Phys.\ Lett.\ B
  \textbf{43} (1973) 403}.

\bibitem{Cashmore:1985vp}
R.~J. Cashmore, C.~M. Hawkes, B.~W. Lynn, and R.~G. Stuart,
  \ifthenelse{\boolean{articletitles}}{\emph{{The forward-backward asymmetry in
  $e^+ e^- \to \mu^+ \mu^-$ comparisons between the theoretical calculations at
  the one loop level in the Standard Model and with the experimental
  measurements}}, }{}\href{https://doi.org/10.1007/BF01560685}{Z.\ Phys.\ C
  \textbf{30} (1986) 125}.

\bibitem{b2tech}
Belle II collaboration, T.~Abe {\em et~al.},
  \ifthenelse{\boolean{articletitles}}{\emph{{Belle II Technical Design
  Report}}, }{}\href{http://arxiv.org/abs/1011.0352}{{\normalfont\ttfamily
  arXiv:1011.0352}}.

\bibitem{Kou:2018nap}
W.~Altmannshofer {\em et~al.}, \ifthenelse{\boolean{articletitles}}{\emph{{The
  Belle~II physics book}}, }{}\href{https://doi.org/10.1093/ptep/ptz106}{PTEP
  \textbf{2019} (2019) 123C01}, Erratum
  \href{https://doi.org/10.1093/ptep/ptaa008}{ibid.\   \textbf{2020} (2020)
  029201}, \href{http://arxiv.org/abs/1808.10567}{{\normalfont\ttfamily
  arXiv:1808.10567}}.

\bibitem{Akai:2018mbz}
K.~Akai, K.~Furukawa, and H.~Koiso,
  \ifthenelse{\boolean{articletitles}}{\emph{{SuperKEKB collider}},
  }{}\href{https://doi.org/10.1016/j.nima.2018.08.017}{Nucl.\ Instrum.\ Meth.\
  A \textbf{907} (2018) 188},
  \href{http://arxiv.org/abs/1809.01958}{{\normalfont\ttfamily
  arXiv:1809.01958}}.

\bibitem{Lange:2001uf}
D.~J. Lange, \ifthenelse{\boolean{articletitles}}{\emph{{The EvtGen particle
  decay simulation package}},
  }{}\href{https://doi.org/10.1016/S0168-9002(01)00089-4}{Nucl.\ Instrum.\
  Meth.\ A \textbf{462} (2001) 152}.

\bibitem{Sjostrand:2014zea}
T.~Sj\"{o}strand {\em et~al.}, \ifthenelse{\boolean{articletitles}}{\emph{{An
  Introduction to PYTHIA 8.2}},
  }{}\href{https://doi.org/10.1016/j.cpc.2015.01.024}{Comput.\ Phys.\ Commun.\
  \textbf{191} (2015) 159},
  \href{http://arxiv.org/abs/1410.3012}{{\normalfont\ttfamily
  arXiv:1410.3012}}.

\bibitem{Jadach:1999vf}
S.~Jadach, B.~F.~L. Ward, and Z.~W\c{a}s,
  \ifthenelse{\boolean{articletitles}}{\emph{{The precision Monte Carlo event
  generator KK for two-fermion final states in $e^+e^-$ collisions}},
  }{}\href{https://doi.org/10.1016/S0010-4655(00)00048-5}{Comput.\ Phys.\
  Commun.\  \textbf{130} (2000) 260},
  \href{http://arxiv.org/abs/hep-ph/9912214}{{\normalfont\ttfamily
  arXiv:hep-ph/9912214}}.

\bibitem{Barberio:1990ms}
E.~Barberio, B.~van Eijk, and Z.~W\c{a}s,
  \ifthenelse{\boolean{articletitles}}{\emph{{PHOTOS: A universal Monte Carlo
  for QED radiative corrections in decays}},
  }{}\href{https://doi.org/10.1016/0010-4655(91)90012-A}{Comput.\ Phys.\
  Commun.\  \textbf{66} (1991) 115}.

\bibitem{Barberio:1993qi}
E.~Barberio and Z.~W\c{a}s, \ifthenelse{\boolean{articletitles}}{\emph{{PHOTOS:
  A Universal Monte Carlo for QED radiative corrections. Version 2.0}},
  }{}\href{https://doi.org/10.1016/0010-4655(94)90074-4}{Comput.\ Phys.\
  Commun.\  \textbf{79} (1994) 291}.

\bibitem{Agostinelli:2002hh}
GEANT4 collaboration, S.~Agostinelli {\em et~al.},
  \ifthenelse{\boolean{articletitles}}{\emph{{GEANT4: A simulation toolkit}},
  }{}\href{https://doi.org/10.1016/S0168-9002(03)01368-8}{Nucl.\ Instrum.\
  Meth.\  \textbf{A506} (2003) 250}.

\bibitem{Kuhr:2018lps}
Belle II Framework Software Group, T.~Kuhr {\em et~al.},
  \ifthenelse{\boolean{articletitles}}{\emph{{The Belle II Core Software}},
  }{}\href{https://doi.org/10.1007/s41781-018-0017-9}{Comput.\ Softw.\ Big
  Sci.\  \textbf{3} (2019) 1},
  \href{http://arxiv.org/abs/1809.04299}{{\normalfont\ttfamily
  arXiv:1809.04299}}.

\bibitem{basf2-zenodo}
{Belle II collaboration}, \ifthenelse{\boolean{articletitles}}{\emph{{Belle II
  Analysis Software Framework (basf2)}}, }{}
  \url{https://doi.org/10.5281/zenodo.5574115}.

\bibitem{PID}
Belle II collaboration, I.~Adachi {\em et~al.},
  \ifthenelse{\boolean{articletitles}}{\emph{{Charged-hadron identification at
  Belle~II}}, }{}\href{http://arxiv.org/abs/2506.04355}{{\normalfont\ttfamily
  arXiv:2506.04355}}.

\bibitem{Longo:2020zqt}
S.~Longo {\em et~al.}, \ifthenelse{\boolean{articletitles}}{\emph{{CsI(Tl)
  pulse shape discrimination with the Belle~II electromagnetic calorimeter as a
  novel method to improve particle identification at electron-positron
  colliders}}, }{}\href{https://doi.org/10.1016/j.nima.2020.164562}{Nucl.\
  Instrum.\ Meth.\ A \textbf{982} (2020) 164562},
  \href{http://arxiv.org/abs/2007.09642}{{\normalfont\ttfamily
  arXiv:2007.09642}}.

\bibitem{Belle-II:2024qod}
Belle and Belle II collaborations, I.~Adachi {\em et~al.},
  \ifthenelse{\boolean{articletitles}}{\emph{{Search for lepton
  flavor-violating decay modes $B^0\to\KS\tau^\pm\ell^\mp$ ($\ell=\mu, e$) with
  hadronic $B$-tagging at Belle and Belle II}},
  }{}\href{http://arxiv.org/abs/2412.16470}{{\normalfont\ttfamily
  arXiv:2412.16470}}.

\bibitem{pdg}
Particle Data Group, S.~Navas {\em et~al.},
  \ifthenelse{\boolean{articletitles}}{\emph{{Review of particle physics}},
  }{}\href{https://doi.org/10.1103/PhysRevD.110.030001}{Phys.\ Rev.\ D
  \textbf{110} (2024) 030001}.

\bibitem{Krohn:2019dlq}
Belle II Analysis Software Group, J.-F. Krohn {\em et~al.},
  \ifthenelse{\boolean{articletitles}}{\emph{{Global decay chain vertex fitting
  at Belle II}}, }{}\href{https://doi.org/10.1016/j.nima.2020.164269}{Nucl.\
  Instrum.\ Meth.\ A \textbf{976} (2020) 164269},
  \href{http://arxiv.org/abs/1901.11198}{{\normalfont\ttfamily
  arXiv:1901.11198}}.

\bibitem{Hocker:2007ht}
H.~Voss, A.~H{\"o}cker, J.~Stelzer, and F.~Tegenfeldt,
  \ifthenelse{\boolean{articletitles}}{\emph{{TMVA, the toolkit for
  multivariate data analysis with ROOT}},
  }{}\href{https://doi.org/10.22323/1.050.0040}{PoS \textbf{ACAT} (2007) 040}.

\bibitem{TMVA2007}
A.~Hoecker {\em et~al.}, \ifthenelse{\boolean{articletitles}}{\emph{{TMVA --
  Toolkit for Multivariate Data Analysis}},
  }{}\href{http://arxiv.org/abs/physics/0703039}{{\normalfont\ttfamily
  arXiv:physics/0703039}}.

\bibitem{Belle-II:2023vra}
Belle~II collaboration, I.~Adachi {\em et~al.},
  \ifthenelse{\boolean{articletitles}}{\emph{{Novel method for the
  identification of the production flavor of neutral charmed mesons}},
  }{}\href{https://doi.org/10.1103/PhysRevD.107.112010}{Phys.\ Rev.\ D
  \textbf{107} (2023) 112010},
  \href{http://arxiv.org/abs/2304.02042}{{\normalfont\ttfamily
  arXiv:2304.02042}}.

\bibitem{johnson}
N.~L. Johnson, \ifthenelse{\boolean{articletitles}}{\emph{{Systems of frequency
  curves generated by methods of translation}},
  }{}\href{https://doi.org/10.1093/biomet/36.1-2.149}{Biometrika \textbf{36}
  (1949) 149}.

\bibitem{LHCb:2014kcb}
LHCb, R.~Aaij {\em et~al.},
  \ifthenelse{\boolean{articletitles}}{\emph{{Measurement of \CP asymmetry in
  $D^0 \rightarrow K^- K^+$ and $D^0 \rightarrow \pi^- \pi^+$ decays}},
  }{}\href{https://doi.org/10.1007/JHEP07(2014)041}{JHEP \textbf{07} (2014)
  041}, \href{http://arxiv.org/abs/1405.2797}{{\normalfont\ttfamily
  arXiv:1405.2797}}.

\end{thebibliography}
\ifthenelse{\boolean{wordcount}}%
{\nobibliography{references}}
{%
\providecommand{\href}[2]{#2}\begingroup\raggedright\endgroup
}

\end{document}